\documentclass[aps,prl,twocolumn,tightenlines,amsmath,amssymb,superscriptaddress]{revtex4-1}

\usepackage{graphicx}
\usepackage{amsmath}
\usepackage{amssymb}
\usepackage{epstopdf}
\usepackage{color}
\usepackage{lipsum}
\usepackage{hyperref}

\DeclareGraphicsExtensions{.eps}

\newcommand{\ket}[1] {|#1 \rangle}

\newcommand*{\rom}[1]{\expandafter\@slowromancap\romannumeral #1@}

\begin{document}

\title{Boson Sampling with single-photon Fock states from a bright solid-state source}

\author{J. C. Loredo}\email[Corresponding author:~]{juan.loredo1@gmail.com}
\affiliation{Centre for Engineered Quantum Systems, Centre for Quantum Computation and Communication Technology, School of Mathematics and Physics, University of Queensland, Brisbane, Queensland 4072, Australia}
\author{M.~A.~Broome}
\affiliation{Centre of Excellence for Quantum Computation and Communication Technology, School of Physics, University of New South Wales, Sydney, New South Wales 2052, Australia}
\author{P. Hilaire}
\affiliation{CNRS-C2N Centre de Nanosciences et de Nanotechnologies, Universit\'e Paris-Sud, Universit\'e Paris-Saclay, 91460 Marcoussis, France}
\affiliation{Universit\'e Paris Diderot-Paris 7, 75205 Paris CEDEX 13, France}
\author{O. Gazzano}
\affiliation{CNRS-C2N Centre de Nanosciences et de Nanotechnologies, Universit\'e Paris-Sud, Universit\'e Paris-Saclay, 91460 Marcoussis, France}
\affiliation{Joint Quantum Institute, National Institute of Standards and Technology,
University of Maryland, Gaithersburg, MD, USA}
\author{I. Sagnes}
\affiliation{CNRS-C2N Centre de Nanosciences et de Nanotechnologies, Universit\'e Paris-Sud, Universit\'e Paris-Saclay, 91460 Marcoussis, France}
\author{A. Lemaitre}
\affiliation{CNRS-C2N Centre de Nanosciences et de Nanotechnologies, Universit\'e Paris-Sud, Universit\'e Paris-Saclay, 91460 Marcoussis, France}
\author{M. P. Almeida}
\affiliation{Centre for Engineered Quantum Systems, Centre for Quantum Computation and Communication Technology, School of Mathematics and Physics, University of Queensland, Brisbane, Queensland 4072, Australia}
\author{P.~Senellart}
\affiliation{CNRS-C2N Centre de Nanosciences et de Nanotechnologies, Universit\'e Paris-Sud, Universit\'e Paris-Saclay, 91460 Marcoussis, France}
\affiliation{D\'epartement de Physique, Ecole Polytechnique, Universit\'e Paris-Saclay, F-91128 Palaiseau, France}
\author{A. G. White}
\affiliation{Centre for Engineered Quantum Systems, Centre for Quantum Computation and Communication Technology, School of Mathematics and Physics, University of Queensland, Brisbane, Queensland 4072, Australia}

\begin{abstract}
{
A \textsc{BosonSampling} device is a quantum machine expected to perform tasks intractable for a classical computer, yet requiring minimal non-classical resources as compared to full-scale quantum computers. Photonic implementations to date employed sources based on inefficient processes that only simulate heralded single-photon statistics when strongly reducing emission probabilities. \textsc{BosonSampling} with only single-photon input has thus never been realised. Here, we report on a \textsc{BosonSampling} device operated with a bright solid-state source of {single-photon Fock states with high photon-number purity}: the emission from an efficient and deterministic quantum dot-micropillar system is demultiplexed into three partially-indistinguishable single-photons, with {a single-photon} purity $1{-}g^{(2)}(0)$ of $0.990{\pm}0.001$, interfering in a linear optics network. Our demultiplexed source is {between one and two orders-of-magnitude more efficient} than current heralded multi-photon sources based on spontaneous parametric downconversion, allowing us to complete the \textsc{BosonSampling} experiment {faster} than previous equivalent implementations.}
\end{abstract}

\maketitle
A core tenet of computer science is the Extended Church-Turing thesis, which states that all computational problems that are efficiently solvable by physically realistic machines are efficiently simulatable with classical resources. In 2011 Aaronson and Arkhipov introduced \textsc{BosonSampling}, a quantum protocol for efficiently sampling the output of a multimode bosonic interferometer~\cite{AA1,AA2,BSIons:Duan,BSth:Saleh,BSVibSpec:Aspuru}: a problem apparently intractable with classical computation. When scaled to many bosons this model of intermediate---i.e.~non-universal---quantum computation will provide the strongest evidence against the Extended Church-Turing thesis.

The most experimentally accessible boson is the photon{, thus serving in the initial experimental implementations of \textsc{BosonSampling}~\cite{broome2013photonic,Spring798,Tillmann:2013kx,Crespi:2013vn,ScatterBS:Sciarrino,UniversalLO:OBrien}.} These {earlier} assays are well short of the numbers of single photons required to probe the Extended Church-Turing thesis: scalable photonic technology is required. The three core technologies needed for scalable quantum photonics are: single-photon sources \cite{Kwiat_1998,pittman2005heralding,babinec2010diamond,yuan2002electrically,santori2002indistinguishable}; large interferometric networks, with current integrated and programmable technology~\cite{intSi:Brien,phaseCont:Walmsley,intPhotGate:Sciarrino,UniversalLO:OBrien}; and efficient photon detection, with demonstrated number resolution~\cite{miller2003demonstration,divochiy2008superconducting}, and efficiencies of up to $95\%$~\cite{Lita08}. 

{To date, \textsc{BosonSampling} implementations employed} photons obtained from spontaneous parametric downconversion, {which output} is far from ideal single-photon Fock states, $\ket{\psi}{=}\ket{1}$, instead producing primarily vacuum with a small admixture of pairs of photons, $\ket{\psi} {=} \sqrt{1{-}|\lambda|^2} \sum_{n=0}^{\infty} \lambda^{n} \ket{nn}$, where $|\lambda| {\ll}1$. A \emph{non-heralded} $2n$-photon source can be built by using $n$ downconverters, but it can only be used in specific protocols where the impact of higher photon-numbers is minimised~\cite{MPQI:Pan}; {alternatively}, it can be operated as a \emph{heralded} $n$-photon source by detecting $n$ photons---one from each downconverter---to herald the presence of their $n$ single-photon partners. Multi-photon rates for state-of-the-art pulsed downconversion sources~\cite{8photon:JWP,4Photon:Walmsley,6PhotonRates:Guo,EQS:Loredo}, pumped at a standard $80$~MHz repetition rate, range from $\sim$300~kHz for $2$ photons---{thus,} yielding heralded single-photons at that rate---down to $\sim$3~mHz for $8$ photons---{accordingly,} $4$ heralded single-photons at that rate. For as little as $6$ heralded single-photons, the rate ($\sim$1 per year) becomes less than the detection rate of gravitational waves~\cite{GWLigo}. 

{Recent progress with time-multiplexing schemes~\cite{timeMux:Kwiat15} can potentially increase these heralded multi-photon rates in future experiments. Using downconversion to manipulate many single-photons remains, however, challenging to date, which has prevented the scaling of \textsc{BosonSampling} to larger photon numbers. In an effort to lessen this hurdle, an extended version of the protocol---named randomized~\cite{BSth:Saleh}, or ``Scattershot''~\cite{ScatterBS:Sciarrino}, \textsc{BosonSampling}---exploits heralding {to obtain an algorithmic enhancement}, by a binomial factor, {in} the number of valid inputs to the protocol: \textsc{BosonSampling} then becomes scalable with probabilistic, but heralded, downconversion sources.}

	\begin{figure*}[htp]
		\includegraphics[width=\textwidth]{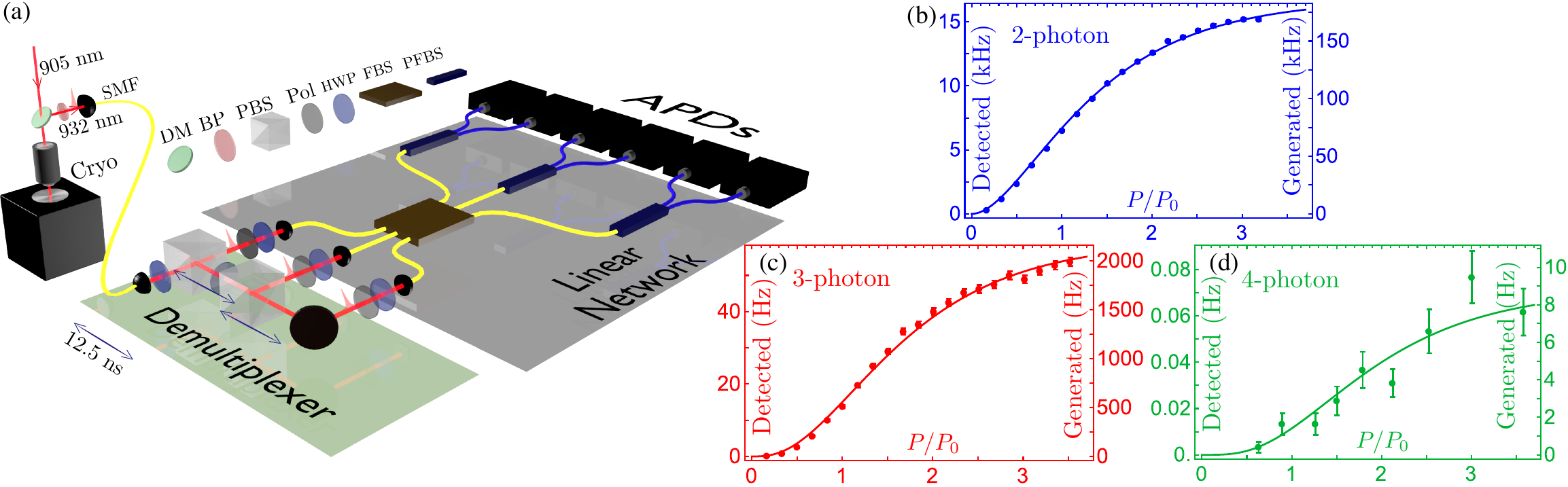}\vspace{-3mm}		
		\caption{{Experimental setup. (a) A dichroic mirror (DM), and a $0.85$~nm FWHM band-pass filter (BP) isolate single-photon emission at $932$~nm from the $905$~nm excitation laser, which is then collected by a single-mode fibre (SMF)}. A passive demultiplexer composed of beam-splitters with tunable transmittances---half-wave plates (HWP), and polarising beam-splitters (PBS)---and compensating delay lines of $12.5$~ns probabilistically converts three consecutive single photons into separate spatial modes {at the input of} the \textsc{BosonSampling} circuit. {The $6{\times}6$ linear network} is composed of polarisers (Pol), half-wave plates, a $3{\times}3$ non-polarising fibre beam-splitter (FBS), and polarising fibre beam-splitters (PFBS). {Six APDs are used to record two- and three-fold correlation measurements to sample from the output distribution of the \textsc{BosonSampling} device. (b)-(d) Detected and generated $n$-photon rates obtained directly from the demultiplexed source. The generated rates include a factor of $(1{/}0.3)^n$ to describe our source modulo detector efficiencies ($30\%$ in average for the used APDs). The $4$-photon count-rates are obtained from the demultiplexer in (a) with an extra tunable beam-splitter. Curves are fits to $c^{(n)}_\text{max}(1{-}e^{-P{/}P_0})^n$, with $c^{(2)}_\text{max}{=}186.4$~kHz, $c^{(3)}_\text{max}{=}2202$~Hz, and $c^{(4)}_\text{max}{=}8.8$~Hz, denoting maximum $n$-photon generated rates.}   }
	\vspace{-4mm}	
	\label{fig:1}
	\end{figure*}
Quantum-dots in photonic structures~\cite{lodahl2004controlling,gazzano2013bright,RevModPhys.87.347,nearIdealSPS:Pan,nearIdealSPS:Senellart} have been recently shown to produce long streams of indistinguishable single-photons with large emission yields~\cite{loredo2016scalable,1000photons:Pan}. Efficient temporal-to-spatial demultiplexing of these sources will enable multi-photon experiments at scales heretofore impossible. Here we implement a \textsc{BosonSampling} device operated with a bright demultiplexed source of three highly-pure single-photon Fock states from the emission of a deterministic quantum dot-micropillar system~\cite{gazzano2013bright}. The high source brightness allows us to {implement multi-photon sources markedly more efficient than their downconversion counterparts, completing the \textsc{BosonSampling} protocol faster than in previous implementations.} Our results prove solid-state sources an appealing candidate to constitute the basis for future quantum photonics, in particular for the implementation of \textsc{BosonSampling} with larger photon numbers.

\noindent {\emph{Source of multiple single-photon Fock states.}}
Laser pulses with a repetition rate of $R_L{=}80$~MHz and wavelength centred at $905$~nm provide quasi-resonant excitation of an InGaAs quantum-dot deterministically coupled to a micropillar cavity, {which itself is housed in an optically accessable cryostat (Cryo) system at $13$~K.} See refs.~\cite{gazzano2013bright,loredo2016scalable} for a detailed description of this quantum dot-micropillar system. An optimised collection efficiency results in a record probability per pump-pulse of finding an spectrally-isolated single-photon at the output of a single-mode fiber---an absolute brightness---of up to $\eta_0{=}0.14$. As a result, our core source generates up to $\sim$11~MHz of single-photons, modulo detector efficiencies, from which $3.6$~MHz are detected with an avalanche photodiode (APD) of $32\%$ quantum efficiency~\cite{loredo2016scalable}. The absolute brightness depends on the laser pump power $P$ according to $\eta{=}\eta_0\left(1{-}e^{-P{/}P_0}\right)$, with $P_0{=}150$~$\mu$W the saturation power. Under quasi-resonant excitation, single-photon sources based on non-gated quantum dots are subject to small and random frequency jitter---known as spectral diffusion---due to charges near the solid-state emitter~\cite{chargeNoise:Warburton,SpectralDiffusion:Kamp}. This results in the emission of photons with partial indistinguishability, which in our case is around $50$--$70\%$ depending on the exact pump conditions~\cite{loredo2016scalable}. {We choose to operate our source at $P{=}1.2P_0$, at which point it exhibits a single-photon purity $1{-}g^{(2)}(0)$ of $0.990{\pm}0.001$, where $g^{(2)}(0){=}0$ holds for an ideal $|n\rangle{=}|1\rangle$ Fock state. Our source remains highly pure even at high pump powers, with a purity of $0.976\pm0.001$ at $3P_0$, see Supplemental Material.}
	
{Temporal to spatial demultiplexing of the source could be achieved with an active---temporally-varying---switcher, such that each of $n$ consecutive single-photons is routed into a different spatial channel, resulting in a scalable method to demultiplex $n$ events from a $1$-photon source into one event of an $n$-photon source.} A simpler alternative is to implement a passive demultiplexer {as depicted in Fig.~\ref{fig:1}({a})}. Here, photon routing occurs by using an array of $n{-}1$ chained beamsplitters with tuned transmittances as to evenly distribute, with probability ${1}{/}{n}$, each single-photon into one of $n$ possible outputs. The high absolute brightness in our core source allows us to readily operate {$2$-, $3$-, and $4$-photon sources with this method. Figures~\ref{fig:1}(b)-(d) show} the detected, and generated---corrected for detector efficiencies---count-rates of our demultiplexed $n$-photon source: $n$ single-photons in the same temporal mode at the output of $n$ single-mode fibres.
	
\begin{figure}[tp]
		\includegraphics[width=.4\textwidth]{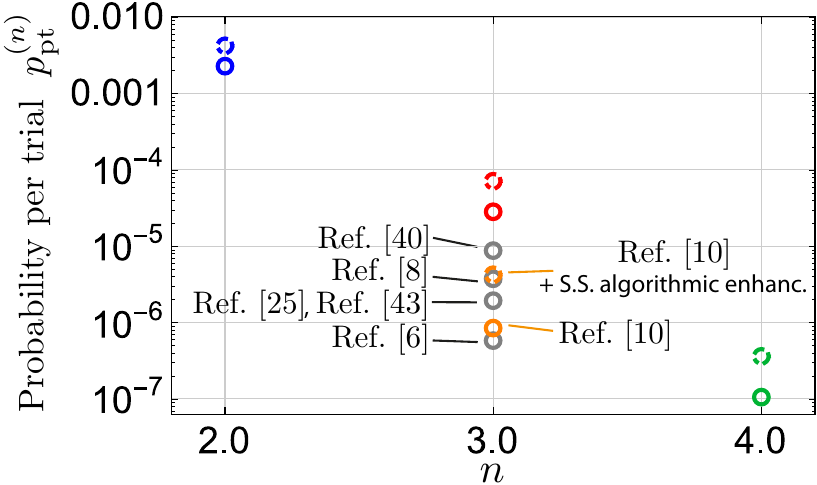}\vspace{-3mm}
		\caption{{Multi-photon source efficiency. $n$-photon probability per trial, $p_{\text{pt}}^{(n)}$, for our 2-, 3-, and 4-photon source taken at $1.2P_0$ (solid {blue, red, and green} circles), and at $3P_0$ (dashed {blue, red, and green} circles). The $p_{\text{pt}}^{(n)}$ is estimated for various downconversion $3$-photon sources (grey and orange circles) employed in previous \textsc{BosonSampling} experiments. {The Scattershot algorithm (S.S.)~\cite{BSth:Saleh,ScatterBS:Sciarrino} results in an effective enhancement of $p_{\text{pt}}^{(n)}$ (orange circles) for its specific protocol.} Our $3$-photon source is between one to two orders-of-magnitude more efficient than the downconversion cases. Note that only partial heralding was employed in all downconversion implementations. A fully heralded $n$-photon source, a necessary condition to produce \emph{true} single-photon Fock state statistics with downconversion, is thus further orders-of-magnitude less efficient than our sources.}}
	\vspace{-0mm}	
	\label{fig:2}
	\end{figure}

{To estimate the efficiency of our source, we define the \emph{$n$-photon probability per trial}, $p_\text{pt}^{(n)}{=}c_\text{gen}^{(n)}{/}R_\text{trial}$, the probability of generating a spectrally-isolated $n$-photon event, at the output of $n$ single-mode fibres, per experimental attempt. Here, $c_\text{gen}^{(n)}$ is the $n$-photon generated rate, and $R_\text{trial}$ is the ``trial'' rate. This allows us to compare multi-photon sources from different systems based solely on their efficiency, irrespective of external parameters, such as detector efficiencies, and pump rates. For an explicit comparison, we compute $p_\text{pt}^{(n)}$ for various \emph{partially} heralded $3$-photon sources used in previous \textsc{BosonSampling} experiments, see Fig.~\ref{fig:2}. Our solid-state based $3$-photon source is {more efficient than} its downconversion counterparts by one to two orders-of-magnitude, see Supplemental Material for details on this comparison. Note that this is achieved using a non-scalable---scaling as $1{/}n^n$---probabilistic demultiplexer. We thus expect our $n$-photon efficiency to increase super-exponentially ($\propto n^n$) with an active demultiplexer.}

	\begin{figure*}[htp]
		\includegraphics[width=.95\textwidth]{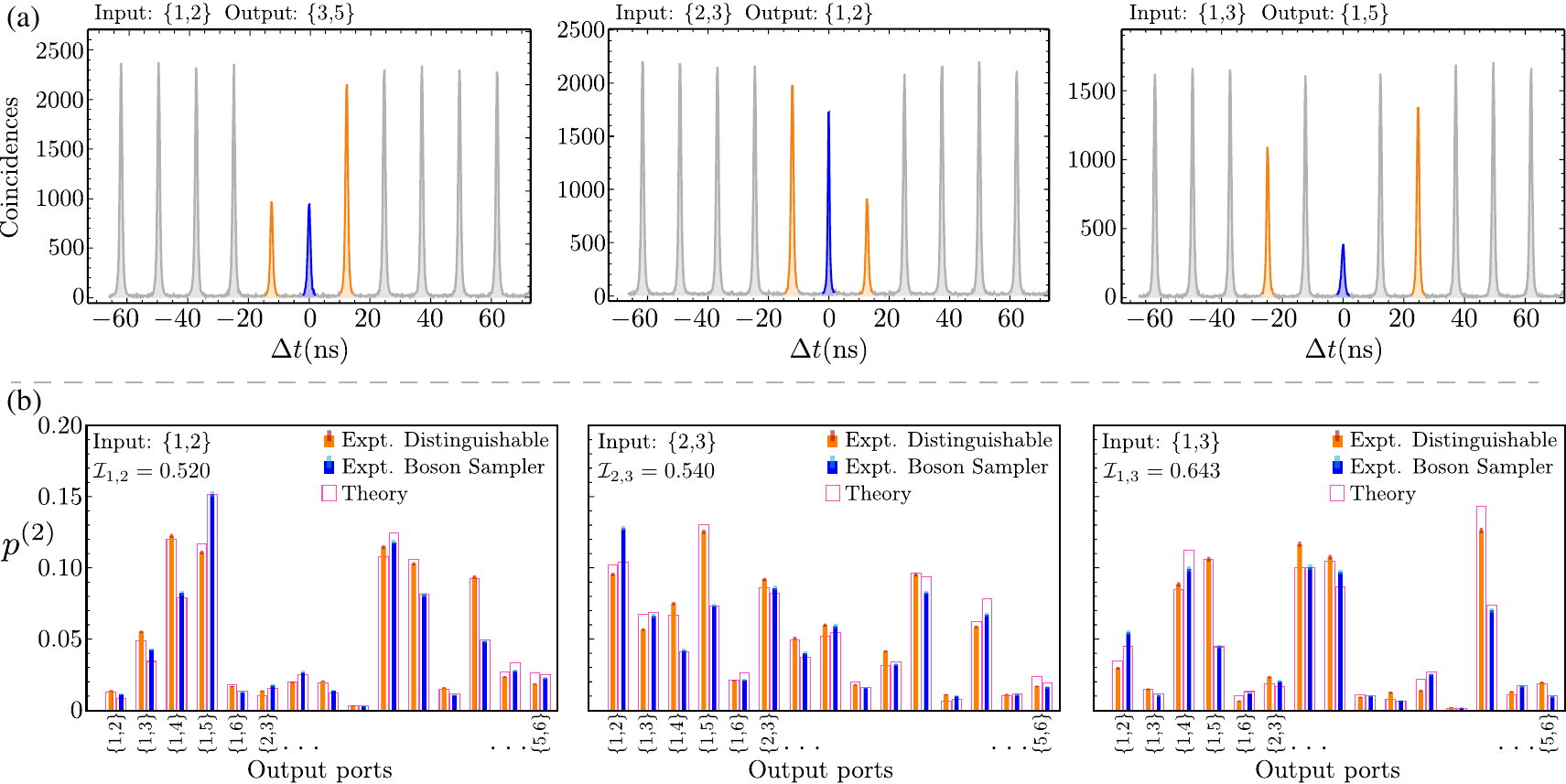}\vspace{-2mm}
		\caption{Two-photon \textsc{BosonSampling}. (a) Temporal-correlation measurements at no-collision outputs for $2$ photons entering at different inputs. Coincidences around $\Delta t{=}0$ (blue peaks) result from two-photon interference and are thus governed by Eq.~(\ref{eq:p2}). The position of reduced areas (orange peaks) indicates the temporal distance in emission from the quantum-dot: For inputs $\{1,2\}$, and $\{2,3\}$, photons were emitted after one laser repetition rate $1{/}R_L{=}12.5$~ns, thus reduced areas appear at $\pm1{/}R_L$; similarly, appearing at $\pm25$~ns for $\{1,3\}$, with photons emitted separated by $2$ laser repetition rates. Coincidences outside $\Delta t{=}0$ (orange peaks and grey peaks) involve non-interfering photons, thus contain only classical information. (b) Coincidences at zero delay from the $15$ no-collision outputs give the distribution of the Boson Sampler (blue bars), with theoretical distributions (empty bars) given by $\mathcal{I}_{1,2}{=}0.520$, $\mathcal{I}_{2,3}{=}0.540$, and $\mathcal{I}_{1,3}{=}0.643$, for their respective input; whereas coincidences outside zero delay determine that of the distinguishable sampler (red bars), with theoretical distribution (empty bars) obtained by assuming zero indistinguishability in Eq.~(\ref{eq:p2}). Note that strong output configurations in the classical sampler \emph{tend} to have a larger reduction when observed in the Boson Sampler. A complete sampled distribution is obtained with $10$~minutes integration time; and, in average, a total of $\sim40000$ $2$-fold events are collected for any given distribution. Error bars (small light-coloured bars) are deduced from assuming poissonian statistics in detected events.}
	\vspace{-2mm}			
	\label{fig:3}
	\end{figure*}	
\noindent {\emph{BosonSampling with solid-state photon sources.}} Using this method, $2$ and $3$ partially-indistinguishable single-photons are used as inputs into the {\textsc{BosonSampling} $6{\times}6$ linear network $\mathcal{L}$,} consisting of $3$ spatial- and $2$ polarisation-encoded modes, see Fig.~\ref{fig:1}(a). The relative temporal delay between photons is fine-tuned as to erase their temporal distinguishability, and the use of polarising fibre beam-splitters ensures that they are indistinguishable in polarisation.
	
We first input $N{=}2$ single-photons, and characterise the $M{=}6$-mode $\mathcal{L}$ network---in general a non-unitary transfer matrix due to inevitable optical losses---using the method described in ref.~\cite{charactLinNetwork:Saleh}, see Supplemental Material. Following the theoretical model developed in ref.~\cite{GenMPI:Walther}, $2$ photons with a degree of indistinguishability quantified by $\mathcal{I}$, entering $\mathcal{L}$ in inputs $\{i,j\}$ and exiting from outputs $\{k_1,k_2\}$ lead to a $2$-fold coincidence probability:
	\begin{equation}\label{eq:p2}
		p^{(2)}=\left(\frac{1+\mathcal{I}}{2}\right)\left|\text{per}(\overline{\mathcal{L}})\right|^2+\left(\frac{1-\mathcal{I}}{2}\right)\left|\text{det}(\overline{\mathcal{L}})\right|^2,
	\end{equation} 
given by the permanent (per) and determinant (det) of the submatrix $\overline{\mathcal{L}}$ formed with rows $i,j$ and columns $k_1,k_2$ of $\mathcal{L}$. Note that Eq.~(\ref{eq:p2}) reduces to the well-known formula $p^{(2)}{=}\left|\text{per}({\overline{\mathcal{L}}})\right|^2$ in the ideal case of perfect indistinguishability, i.e. $\mathcal{I}{=}1$.

We measured all $\binom{M}{N}{=}15$ outputs in which photons exit $\mathcal{L}$ in different modes, so-called no-collision events. Peak areas in temporal-correlation measurements at these outputs allow us to extract---in a single experimental run---both the sampling distribution resulting from the Boson Sampler---that is, with partially-indistinguishable photons---and that of a (classical) distinguishable sampler arising from completely distinguishable particles. Given an output configuration $k$, coincidences detected under the area $A^{0}_k$ around zero delay $\Delta t{=}0$ are subject to two-photon interference: they determine the Boson Sampler distribution by measuring $\overline p^{(2)}_k{=}A^{0}_k$. Conversely, photons leading to coincidences around $\Delta t{=}\pm{l}\times$(12.5~ns), for $l$ integer, do not interfere, and one would expect that these distributions contain information of a classical sampler. Indeed, following ref.~\cite{loredo2016scalable}, one can deduce that the distinguishable sampler distribution is measured via $\overline p^{(2)}_k(0){=}2A^{r}_k{-}A^{n}_k{-}A^{p}_k$, where $A^{r}_k$ is a reference area (average in grey peaks),  $A^{n}_k$ is the reduced area at negative $\Delta t$ (left orange peak), and $A^{p}_k$ is the reduced area at positive $\Delta t$ (right orange peak) as shown in Fig.~\ref{fig:3}(a). Measuring only no-collision events, however, does not provide access to the entire output distribution, thus to obtain probabilities we normalise the measured distributions to the corresponding theoretical prediction according to Eq.~(\ref{eq:p2})---that is, the sum of experimentally obtained probabilities within the no-collision subspace is matched to that as in theory; and, given a $2$-photon input $\{i,j\}$, $\mathcal{I}_{i,j}$ is extracted from the measured output distribution, see Supplemental Material. 

Figure~\ref{fig:3}(b) shows our $2$-photon \textsc{BosonSampling} results. Experimental distributions for the Boson Sampler (blue bars) are shown for $3$ different $2$-photon inputs, and their theoretical distributions (empty bars) are obtained with pair-wise indistinguishabilities $\mathcal{I}_{1,2}{=}0.520$, $\mathcal{I}_{2,3}{=}0.540$, and $\mathcal{I}_{1,3}{=}0.643$, respectively; in agreement with independently measured indistinguishabilities via two-photon interference on a $2\times2$ beamsplitter, see Supplemental Material. For the distinguishable sampler (red bars), the theoretical distribution (empty bars) is calculated by using $\mathcal{I}_{i,j}{=}0$, $\forall i,j$ in Eq.~(\ref{eq:p2}). To quantify the agreement between theory and experiment, we employ the statistical fidelity $\mathcal{F}{=}\sum_i\sqrt{p^{th}_{i}p^{exp}_{i}}$ between normalised theoretical and experimental distributions. For our $2$-photon \textsc{BosonSampling}, we find an average fidelity of $\overline{\mathcal{F}}{=}0.9984{\pm}0.0007$ across the six sampled distributions in Fig.~\ref{fig:3}(b), where the error here is one standard deviation among the six fidelity values.

We now tune the source to input $N{=}3$ single-photons into the $\{1,2,3\}$ mode. In this case, the probability of detecting a $3$-fold coincidence at outputs of $\mathcal{L}$ is~\cite{GenMPI:Walther}: 
\vspace{-1mm}
\begin{equation}\label{eq:p3}
		p^{(3)}=t_6^\dagger\left( \mathbb{I}+\sum_{i\neq j}\rho_{i,j}\mathcal{I}_{i,j}+{\tilde\rho}\prod_{i\neq j}{\sqrt{\mathcal{I}_{i,j}}} \right)t_6,
\end{equation}
with $\mathbb{I}$, the $6{\times}6$ identity operator; $t_6$, a $6$-component quantity that depends on the permanent, determinant, and immanants of $3\times3$ submatrices $\mathcal{T}$; and the $\rho_{i,j}$, and ${\tilde{\rho}}$ matrices as explicitly defined in the Supplemental Material. Eq.~(\ref{eq:p3}) reduces to $p^{(3)}{=}\left|\text{per}(\mathcal{T})\right|^2$ in the ideal case of perfect indistinguishability between all particles, i.e. $\mathcal{I}_{i,j}{=}1$, $\forall i,j$.

	\begin{figure*}[htp]
		\includegraphics[width=.95\textwidth]{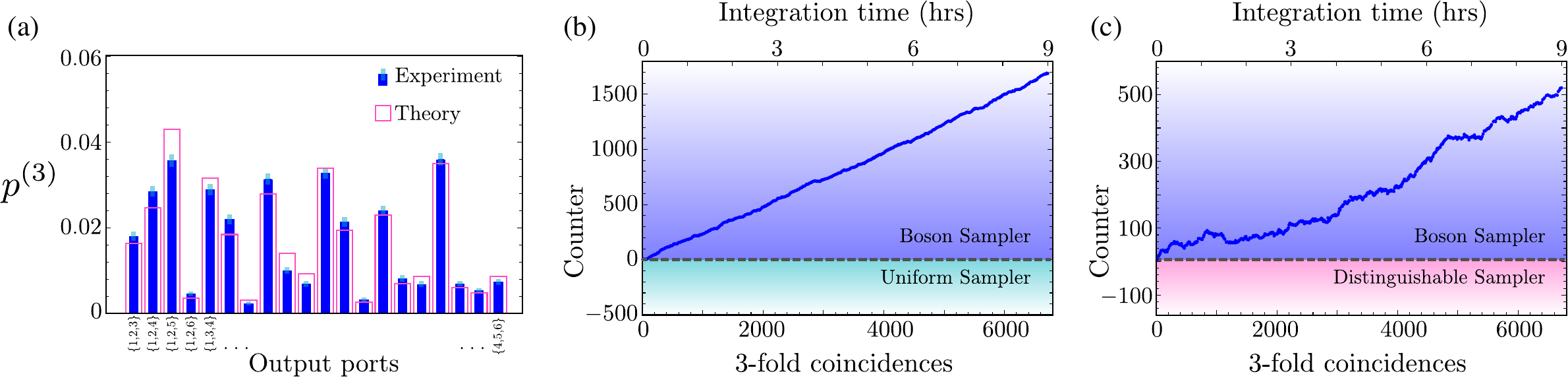}\vspace{-2mm}
		\caption{Three-photon \textsc{BosonSampling}. (a) A total of $20$ no-collision $3$-fold simultaneous coincidences are recorded to obtain the Boson Sampler distribution (blue bars); the theoretical distribution (empty bars) is obtained from Eq.~(\ref{eq:p3}) and by using the previously determined pair-wise indistinguishability parameters. Error bars (light-coloured bars) are deduced from poissonian statistics in measured events. We apply the validation of \textsc{BosonSampling} protocol against the uniform sampler (b), and distinguishable sampler (c). A counter (blue dots) is updated for every $3$-fold event and at any point a positive value validates the data as being obtained from a Boson Sampler as opposed to either a uniform or distinguishable sampler, see Supplemental Material. The final data set contains a total of $6725$ $3$-fold events collected in $9$ hours, that is $\sim$1000 per $80$ minutes; {a faster rate than in previous \textsc{BosonSampling} experiments.}} 
	\vspace{-4mm}			
	\label{fig:4}
	\end{figure*}	
Verifying the output distribution of a \textsc{BosonSampling} device involves calculating a number of (modulus squared) matrix permanents. This task is in general computationally hard to implement efficiently on a classical computer. The complete result of a large-scale \textsc{BosonSampling} machine is thus likely to be, even in principal, unverifiable. It has been even argued that a large-scale \textsc{BosonSampling} experiment will fail to distinguish its data from the (trivial) uniform distribution---i.e., one in which every output configuration is equally probable~\cite{UnifDist:Eisert}. In light of this, some methods have been proposed and demonstrated for the \emph{validation} of \textsc{BosonSampling}: circumstantial evidence is provided to support that a \textsc{BosonSampling} machine is indeed functioning according to the laws of quantum mechanics, by ruling out that the experimentally obtained data originates from, e.g., the uniform distribution, or a sampler with distinguishable particles~\cite{AAValidation,BSValidate:Sciarrino,BSValidate:Obrien,UniversalLO:OBrien}.

Figure~\ref{fig:4} shows our experimental results for the $3$-photon Boson Sampler. In Fig.~\ref{fig:4}(a), the previously determined $2$-photon indistinguishabilities $\mathcal{I}_{i,j}$ are used as input for the theoretical distribution (empty bars) according to Eq.~(\ref{eq:p3}), and experimental probabilities (blue bars) are obtained by measuring the $\binom{M}{N}{=}20$ $3$-fold simultaneous---i.e. around $\Delta t{=}0$---coincidences for no-collision events normalised to the theoretical prediction. We find the $3$-photon \textsc{BosonSampling} fidelity ${\mathcal{F}}{=}0.997{\pm}0.006$, where the error here results from propagated poissonian statistics. In Figs.~\ref{fig:4}(b),(c), we apply the validation of \textsc{BosonSampling} protocol to our data. We record $3$-fold coincidences in steps of $30$ seconds, in which time a counter is updated. For each detected $3$-fold coincidence, the counter is either increased or decreased in one unit, and it is designed, see Supplemental Material, such that after an experimental run a positive value validates the data as obtained from the Boson Sampler distribution, whereas a negative counter indicates it originates from the uniform sampler, see Fig.~\ref{fig:4}(b), or the distinguishable sampler, see Fig.~\ref{fig:4}(c). We observed overall increasing positive counters, thus validating our \textsc{BosonSampling} device by ruling out the alternative hypotheses.

{Note that aside these validation protocols, the increasing interest in resolving the quantum or classical nature of, in general, quantum optical experiments has recently resulted in more general approaches to identify when a device can be efficiently simulated by classical means~\cite{SuffCond:Saleh16}.}

\noindent {\emph{Discussion}}
We experimentally demonstrated multi-photon interference with a highly-efficient solid-state source: a \textsc{BosonSampling} device implemented with single-photon Fock states emitted by a deterministic quantum dot-micropillar system. {A temporal to spatial demultiplexing scheme resulted in multi-photon sources between one to two orders-of-magnitude more efficient than their downconversion versions, which allowed us to complete the \textsc{BosonSampling} protocol faster than in previous experiments~\cite{broome2013photonic,Spring798,Tillmann:2013kx,Crespi:2013vn}. An active source demultiplexing would further boost our multi-photon efficiency super-exponentially---with the number of photons---potentially enabling \textsc{BosonSampling} with larger photon numbers.}

{Furthermore}, we directly observed the effect of partial distinguishability: Our results follow closely the sampling of permanents and immanants of matrices with contributions {modulated by photon} indistinguishability. Moreover, by exploiting temporal-correlation measurements we showed that both classical and quantum $2$-photon sampling distributions can be obtained simultaneously, which can be readily extended to multi-fold temporal-dependent measurements in a larger \textsc{BosonSampling} experiment. Potentially, this could motivate new validation protocols exploiting statistics that include this temporal degree of freedom.

The impact of partial distinguishability in \textsc{BosonSampling} has been studied theoretically~\cite{BSTheory:Shchesnovich,BSTheory:Tichy,GenMPI:Walther,MBCS:Tamma}, and reported experimentally~\cite{GenMPI:Walther}. However, identifying experimentally this property in isolation is challenging. Previous experiments with downconversion exhibit photon-statistics polluted by higher-order terms~\cite{MPQI:Pan}, which can be mistakenly interpreted as decreased photon-indistinguishability. In fact, in many cases these higher-order terms, and not photon distinguishability, are the main cause of performance degradation in downconversion-based protocols~\cite{HiherOrder:Till,Marco2009}. The pathway to maximise indistinguishability in efficient solid-state sources is well known: resonant excitation of the quantum-dot results in near-optimal values of photon indistinguishability~\cite{nearIdealSPS:Pan,nearIdealSPS:Senellart}, in which case the obtained output distributions will be close to the sampling of only permanents---functions belonging to the $\#$P complexity class, in which the main complexity arguments of \textsc{BosonSampling} apply.

We believe our results pave the way to the forthcoming advent of quantum-dot based quantum photonics, in which a future \textsc{BosonSampling} implementation with efficiently demultiplexed and resonantly-pumped solid-state sources may finally see the Extended Church-Turing thesis put to serious test.

This work was partially supported by the Centre for Engineered Quantum Systems (Grant No. CE110001013), the Centre for Quantum Computation and Communication Technology (Grant No. CE110001027), the Asian Office of Aerospace Research and Development (grant FA2386-13-1-4070), by the ERC Starting Grant No. 277885 QD-CQED, the French Agence Nationale pour la Recherche (ANR DELIGHT, ANR USSEPP), the French RENATECH network, the Labex NanoSaclay. A.G.W. acknowledges support from a UQ Vice-Chancellor's Research and Teaching Fellowship. J.C.L., M.P.A., and A.G.W.  thank Devon Biggerstaff for experimental assistance, and the team from the Austrian Institute of Technology for kindly providing time-tagging modules. J.C.L. thanks Saleh Rahimi-Keshari, Marco Bentivegna, Fabio Sciarrino{, and Max Tillmann} for valuable discussions.

{After this Letter was submitted, we became aware of related works~\cite{BSQDot:Pan16,BSQDot:Pan17}.}

\bibliography{Boson_Sampling}

\section{Supplemental Material}

\subsection{I.~Single-photon purity}
\begin{figure*}[htp!]
		\begin{center}
		\includegraphics [width=14.cm]{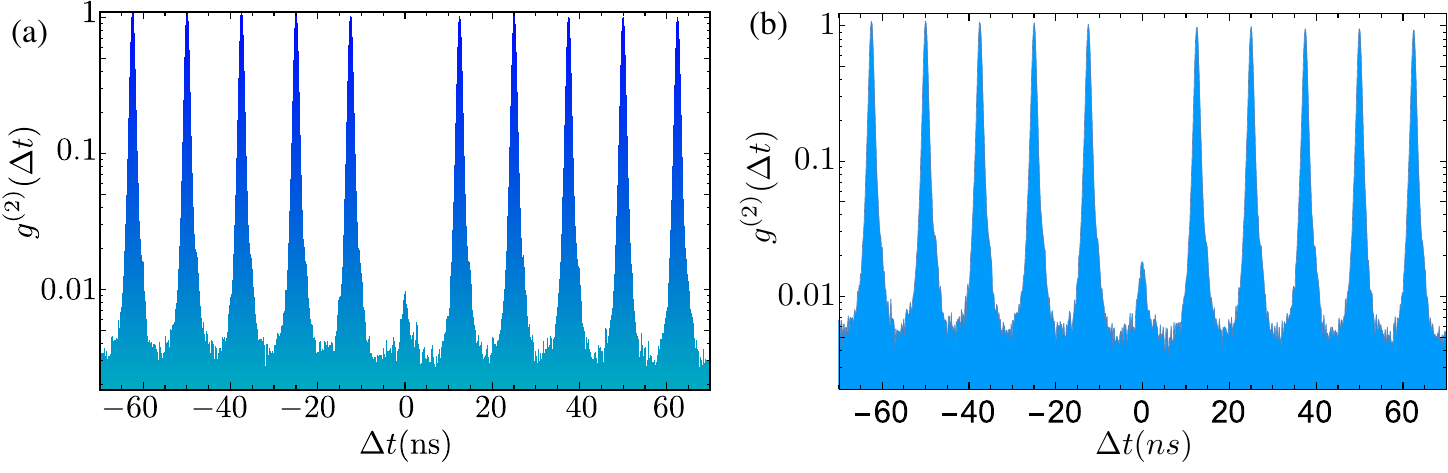}\vspace{-4mm}
		\end{center}
\caption{Second-order autocorrelation function $g^{(2)}(\Delta t)$ (log scale). A Hanbury Brown and Twiss experiment results in coincidences every $1{/}R_L{=}12.5$~ns. Decreased detected events---antibunching---at $\Delta t{=}0$ indicates non-classical states of light, where an ideal single-photon Fock state exhibits $g^{(2)}(0){=}0$. We measure (a) $g^{(2)}(0){=}0.010\pm0.001$ at $P{=}1.2P_0$, and (b) $g^{(2)}(0){=}0.024\pm0.001$ at $P{=}3P_0$, resulting in single-photon purities $1{-}g^{(2)}(0)$ of $0.990\pm0.001$ and $0.976\pm0.001$, respectively.}
\label{fig:SM_BS1}\vspace{-2mm}
	\end{figure*}

Figure~\ref{fig:SM_BS1} shows the single-photon purity of our source from autocorrelation measurements at $1.2$, and $3$ times the saturation power $P_0$.

\subsection{II.~$n$-photon probability per trial}

In the main text, the \emph{$n$-photon probability per trial}, $p_\text{pt}^{(n)}$, is defined as the probability of generating a spectrally-isolated n-photon event, at the output of n single-mode fibres, per experimental attempt. Here, we expand on this concept, and elaborate on what we consider as a ``trial'', or ``experimental attempt''.

First, a relevant concept of $n$-photon efficiency is that taken at a point in which the source is readily useful, for which reason we consider $n$-photon events after all spectral filtering needed to perform the experiment; at the output of single-mode fibres as to straightforwardly interconnect it with a given protocol setup; and only the probability of \emph{generating} the event---corrected for detector efficiencies---is considered to be insensitive to different detector performances at different operating wavelengths.

Secondly, different sources---either from the same, or from different systems---involve distinct experimental attempts to generate them. For instance, a $3$-photon source can be obtained from:

\noindent 1) A second-order downconversion event generated from 1 single laser pulse, this generates two pairs of photons, one of which is used to partially herald the source.\\
\noindent 2) First-order downconversion events from 2 laser pulses in a double-pass pump configuration, which generates one pair in the forward direction, one pair in the backward direction, and one of the four photons is used to partially herald the source.\\
\noindent 3) Consecutive single-photon emission from a quantum dot generated after 3 laser pulses, as it is the case in our experiment.

	\begin{table}[!t]
	\centering
		\begin{tabular}{|c c c c c|} 
 \hline
~~$\text{Ref.}$~~&~~$c_\text{det}^{(3)}$~(Hz)~~&~~$\eta_\text{d}$~~&~~$R_\text{trial} (\text{Hz})$~~&~~$p_\text{pt}^{(3)}$~~ \\  [1ex] 
 \hline\hline
 \text{This Work }$(3P_0)$ & $51$ & $(0.3)^3$ & $2.7{\times}10^7$ & $7.1{\times}10^{-5}$ \\ 
 \text{This Work }$(1.2P_0)$ & $20$ & $(0.3)^3$ & $2.7{\times}10^7$ & $2.8{\times}10^{-5}$ \\  
 \cite{GenMPI:Walther} & $91$ & $(0.6)^4$ & $8{\times}10^7$ & $8.7{\times}10^{-6}$ \\ 
 \cite{Tillmann:2013kx} & $39$ & $(0.6)^4$ & $8{\times}10^7$ & $3.7{\times}10^{-6}$ \\
 \cite{4Photon:Walmsley} & $20$ & $(0.6)^4$ & $8{\times}10^7$ & $1.9{\times}10^{-6}$ \\
 \cite{BSValidate:Sciarrino} & $20$ & $(0.6)^4$ & $8{\times}10^7$ & $1.9{\times}10^{-6}$ \\
 \cite{ScatterBS:Sciarrino} & $9$ & $(0.6)^4$ & $8{\times}10^7$ & $8.4{\times}10^{-7}$ \\ 
 \cite{ScatterBS:Sciarrino} $+$ Scattershot  & $45$ & $(0.6)^4$ & $8{\times}10^7$ & $4.2{\times}10^{-6}$ \\ 
 \cite{broome2013photonic} & $6$ & $(0.6)^4$ & $8{\times}10^7$ & $5.8{\times}10^{-7}$ \\ [1ex] 
 \hline
		\end{tabular}
	\vspace{-1mm}		
		\caption{{\bf 3-photon source efficiency.} Parameters used in estimating $p_\text{pt}^{(3)}$. For our work, we measured an average detector efficiency of 0.3, and three detectors were used. For the other references, we assumed a detector efficiency of 0.6, the expected value at downconversion wavelengths, and four detectors (three plus heralding) were  used.}
	\vspace{-5mm}		
	\label{tableSM:counts3}
	\end{table}
In the above examples, distinct approaches will lead to a different amount of attempts per unit of time to generate a $3$-photon event. Assuming a pulsed laser with a standard 80~MHz repetition rate:  For 1), we attempt to produce the source 8${\times}10^7$ times a second. For 2), although twice the number of pulses per second are sent into a non-linear crystal, we still attempt 8${\times}10^7$ times a second to generate the source. In 3), one needs 3 pulses to generate the state, thus the number of attempts per second is reduced to $\sim2.7{\times}10^7$.

Taken this into account, we can calculate $p_\text{pt}^{(n)}$:
	\begin{equation}
		p_\text{pt}^{(n)}=\frac{c_\text{gen}^{(n)}}{R_\text{trial}}=\frac{c_\text{det}^{(n)}}{\eta_\text{d}R_\text{trial}},
	\end{equation}
where $c_\text{det}^{(n)}$ ($c_\text{gen}^{(n)}$) is the detected (generated) $n$-photon rate; $n_\text{d}$ is the total efficiency accounting for all detectors employed, e.g., a non-heralded downconversion $n$-photon source uses $n$ detectors, whereas a fully heralded one uses $2n$; and $R_\text{trial}$ is the rate of trials.

Table~\ref{tableSM:counts3} summarizes the specific values employed to calculate the 3-photon $p_\text{pt}^{(3)}$, which was used in the comparison between our source and those used in previous \textsc{BosonSampling} experiments with downconversion. The detected rates $c_\text{det}^{(3)}$ used in refs.~\cite{GenMPI:Walther,Tillmann:2013kx} were obtained via private communication, and rates of $90.5$~Hz, and $38.7$~Hz were provided. For refs~\cite{4Photon:Walmsley,BSValidate:Sciarrino}, the $20$~Hz 4-photon rates (3-photon plus heralding) were obtained from the manuscripts. For ref.~\cite{ScatterBS:Sciarrino}, 35~kHz, and 20~kHz 2-photon rates are reported in the Supplementary Materials, from where a 4-photon rate (3-photon plus heralding) of $35~\text{kHz}*20~\text{kHz}{/}(80~\text{MHz})=8.75~\text{Hz}$ is derived. The Scattershot approach results in an increase---in this case, a binomial factor of $\binom{5}{1}{=}5$---in the number of valid inputs, effectively increasing $p_\text{pt}^{(n)}$ for the protocol.
For ref.~\cite{broome2013photonic}, a detected 4-photon rate of 1.2~kHz is reported at 100\% pump power, which after spectral filtering of 3 photons (measured filter transmission of 0.5), and 20\% pump power operation, is reduced to a 4-photon rate (3-photon plus heralding) of 6~Hz. The values reported for our sources are extracted from the power dependent $n$-photon saturation curves, and we employed the parameters $\eta_\text{d}{=}(0.3)^n$, and $R_\text{trial}{=}8{\times}10^7{/}n$~Hz for estimating $p_\text{pt}^{(n)}$.

\subsection{III.~Expected rates}
The expected $n$-photon count-rate is:
	\begin{equation}\label{figSMBS:cn}
		c^{(n)}{=}\left(\eta_0\left(1-e^{-P/P_0}\right)\eta_\text{setup}\right)^n\left(\frac{1}{n}\right)^nR_L,
	\end{equation}
where $\eta_0{=}0.14$ is the measured maximum absolute brightness, $\eta_0\left(1-e^{-P/P_0}\right)$ is the absolute brightness at a given relative pump power $P{/}P_0$, $\eta_\text{setup}$ accounts for the experimental setup transmission and detection efficiencies, the factor $\left({1}{/}{n}\right)^n$ is due to the probabilistic nature of the demultiplexer, and $R_L$ is the laser's repetition rate.

We operate our source at $R_L{=}80$~MHz.~The measured optical transmission of our demultiplexer is $\eta_\text{demux}{=}0.650$, arising from 3 polarizing beam-splitters, 15 AR-coated mirrors, and single-mode fibre couplers; which together with an average detector efficiency of $\eta_\text{det}{=}0.30$ results in a setup efficiency of $\eta_\text{setup}{=}\eta_\text{demux}\eta_\text{det}{=}0.195$. At $P{/}P_0{=}3$, these parameters {predict}, according to Eq.~(\ref{figSMBS:cn}), detecting count-rates of $c^{(2)}{=}13.5$~kHz, $c^{(3)}{=}52$~Hz, and $c^{(4)}{=}0.14$~Hz, in good agreement with the actual detected count-rates $c^{(2)}_\text{det}{=}15.1$~kHz, $c^{(3)}_\text{det}{=}51$~Hz, and a discrepancy to the measured $c^{(4)}_\text{det}{=}0.06$~Hz can be attributed to a relatively large measurement error, see Fig.~1 of the main text.

Our \textsc{BosonSampling} setup contains a free-space preparation stage with $\eta_\text{prep}{=}0.723$, an average coupling into single-modes of a $3{\times}3$ fibre beam-splitter of $\eta_\text{fc}{=}0.877$, transmission of such fibre beam-splitter of $\eta_\text{fbs}{=}0.678$, and an average transmission of polarizing fibre beam-splitter of $\eta_\text{pfbs}{=}0.767$. This results in a combined \textsc{BosonSampling} setup efficiency of $\eta_\text{setup}^\text{BS}{=}\eta_\text{demux}\eta_\text{prep}\eta_\text{fc}\eta_\text{fbs}\eta_\text{pfbs}\eta_\text{det}{=}0.064$; which at $P{/}P_0{=}1.2$, according to Eq.~(\ref{figSMBS:cn}), predicts $c^{(3)}{=}0.73$~Hz, the total $3$-fold count-rate that we would expect with completely distinguishable particles fed into the \textsc{BosonSampling} experiment. This is consistent with our measured total 3-fold count-rate of $0.21$~Hz (6725 3-fold events collected in 9 hours) in an experiment performed with partially-indistinguishable particles.	
	
\subsection{IV.~Transfer matrix}
	\begin{figure*}[htp!]
		\begin{center}
		\includegraphics [width=.8\textwidth]{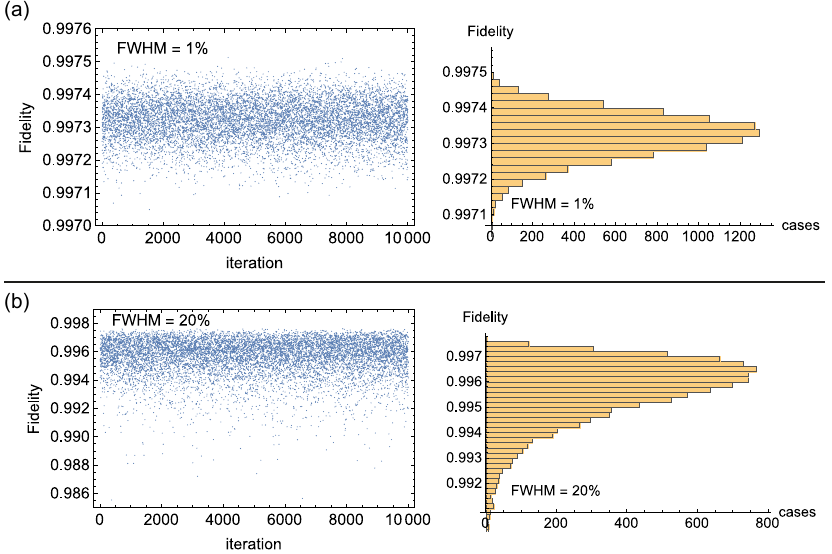}\vspace{-4mm}
		\end{center}
\caption{Impact of detectors' efficiency mismatch. Statistical fidelity of 10000 iterations for (a) FWHM${=}1\%$ (left) and (b) FWHM${=}20\%$ (left), and histograms (right) showing the number of cases that fall within a given fidelity bin.}
\label{fig:SM_FigFidelities}\vspace{-2mm}
	\end{figure*}
{The linear network is composed by a $3{\times}3$ fibre beam-splitter, defining 3 spatial modes; and 3 polarizing fibre beam-splitters, giving access to 2 polarization-encoded modes; which combined result in a $6{\times}6$ network. Stress applied on these fibres before the experiment tunes a network $\mathcal{L}$ to an unknown configuration, which is then characterized with the method introduced in ref.~\cite{charactLinNetwork:Saleh}.} {This method consists of measuring: the probability $|\mathcal{L}_{i,j}|^2$  of a photon entering $\mathcal{L}$ in input $i$ and exiting in output $j$, and phase factors $\arg(\mathcal{L}_{i,j})$ obtained from classical interference patterns. These measurements allow to reconstruct the complex elements $\mathcal{L}_{i,j}$.} For all measurements presented in the main text, inputs $1$, $2$, and $3$ of $\mathcal{L}$ are used. The transfer matrix $\mathcal{L}$ in this subspace is given by:
	\begin{widetext}
		\begin{equation}\label{eqSM:lexp}
			\mathcal{L}=\begin{bmatrix}
0.314 & 0.160 & 0.251 & 0.578 & 0.576 & 0.188 \\ 
0.561 & -0.157+0.151i & -0.319+0.440i & -0.388-0.033i & 0.331-0.127i & -0.120-0.226i\\ 
0.473 & 0.352+0.409i & -0.054-0.025i & 0.249-0.206i & -0.559+0.112i & 0.085-0.118i
			\end{bmatrix}.
		\end{equation}
	\end{widetext}
{Measurement errors arise primarily from obtaining $|\mathcal{L}_{i,j}|^2$, due to power instabilities of the laser light used for the characterization. The \emph{relative} errors in these measurements are all ${<}0.01$, with an average value of 0.007.}

The obtained $\mathcal{L}$, as in Eq.~(\ref{eqSM:lexp}), is then used to calculate $p^{(n)}$, see main text. In practice, the experimentally obtained output distribution can be slightly biased away from the theoretical prediction due to being obtained with various single-photon detectors (APDs) with different efficiencies---as opposed to the classical reconstruction of $\mathcal{L}$, where only one photodiode was used---what, in turn, will affect protocol fidelities. We model the effect of different detectors' efficiencies in final protocol fidelities: $p^{(3)}$, for instance, is multiplied by three relative efficiencies---belonging to the corresponding detectors of a given output---whose values are given by a random variable normally distributed around unity.

We iterated this simulation 10000 times, and computed the statistical fidelity, see Fig.~\ref{fig:SM_FigFidelities}, to the theoretical prediction of $p^{(3)}$. Figure~\ref{fig:SM_FigFidelities}a shows a case where the normal distribution has a small full-width half-maximum (FWHM) of $1\%$. As expected, the resulting fidelities are distributed closely around the reported fidelity $\mathcal{F}{=}0.997$---value rounded to $10^{-3}$ precision---as in the main text. Figure~\ref{fig:SM_FigFidelities}b illustrates a case with FWHM${=}20\%$: the statistical fidelity is not largely affected even in this case with larger variation. In our experiment, we measured the relative efficiencies of our 6 APDs to be $0.98, 1.00, 1.02, 0.95, 0.93, 0.97$; from which we obtain a distribution with a fidelity of $\mathcal{F}{=}0.9998$ to the case with uniform efficiencies---thus having a minimal impact in our measurements.

	\begin{figure*}[htp]
		\begin{center}
		\includegraphics [width= 17.cm]{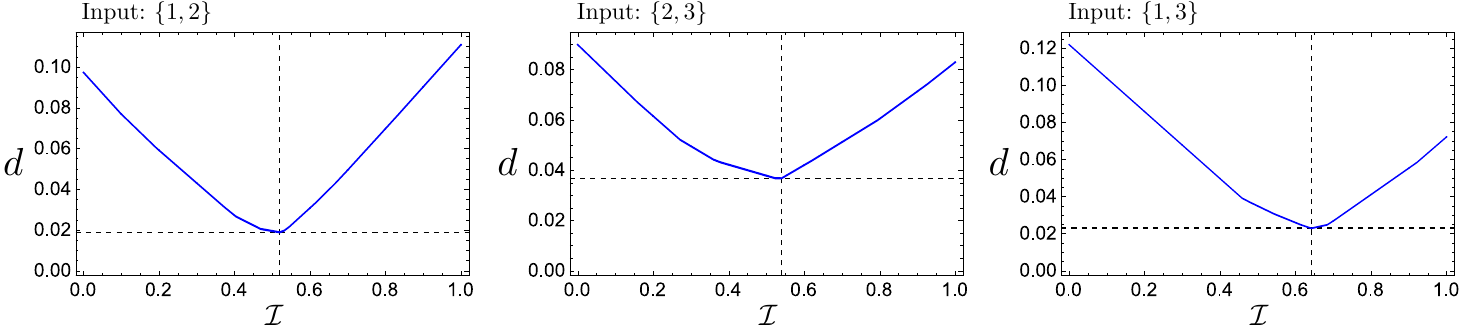}
		\end{center}
\caption{Variation distance $d$ between experimental and theoretical distributions. $d$ is a global minimum at $\mathcal{I}_{1,2}{=}0.520$, $\mathcal{I}_{2,3}{=}0.540$, and $\mathcal{I}_{1,3}{=}0.643$ for its corresponding $2$-photon input. The variation distances at these points are respectively $d_{1,2}{=}0.019$, $d_{2,3}{=}0.037$, and $d_{1,3}{=}0.023$. These values are obtained with both experimental and theoretical distributions normalised to the non-unity theoretical normalisation factor. When the distributions are normalised to unity, the variation distances are $d_{1,2}{=}0.028$, $d_{2,3}{=}0.049$, and $d_{1,3}{=}0.055$, respectively.}
\label{fig:SM_BS2}
	\end{figure*}
		\begin{figure*}[htp]
		\begin{center}
		\includegraphics [width= 15.cm]{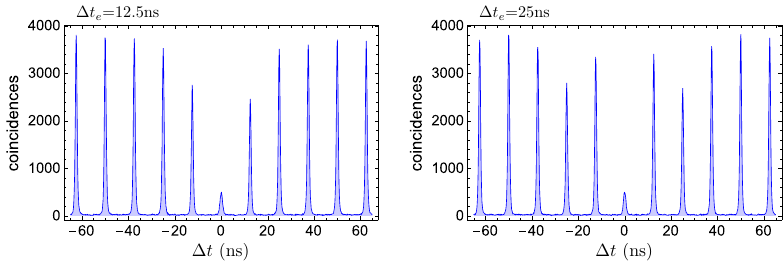}
		\end{center}
\caption{Two-photon interference on a $2\times2$ beam-splitter. Temporal-correlation measurements result in a series of peaks from which the degree of indistinguishability can be directly extracted via $\mathcal{I}{=}(R^2{+}T^2-A_0{/}A)/(2RT)$, with $R{=}0.471$ the beam-splitter reflectance, $T{=}1{-}R$, $A$ the average peak area outside $\Delta t{=}0$ (excluding reduced peaks at $\Delta t_e$), and $A_0$ the peak area around $\Delta t{=}0$. See ref.~\cite{loredo2016scalable} for a derivation of this formula. We obtain $\mathcal{I}^{bs}_{12.5\text{ns}}{=}0.6360{\pm}0.0063$ for $\Delta t_e{=}12.5$~ns, and $\mathcal{I}^{bs}_{25\text{ns}}{=}0.6252{\pm}0.0065$ for $\Delta t_e{=}25$~ns. Errors are estimated from propagated poissonian statistics.}
\label{fig:SM_BS3}
	\end{figure*}
\subsection{V.~Pair-wise indistinguishability}
{In a \textsc{BosonSampling} experiment, the main parameters changing the output distribution are particle distinguishability---originating from either spectral, spatial, or temporal mismatch---and higher-order photon terms. The high single-photon purity of our source, as shown in Fig.~\ref{fig:SM_BS1}, evidences that higher-order terms have a negligible impact. The major parameter that modulates the output of our experiment is thus the pair-wise photon indistinguishability $\mathcal{I}_{i,j}$ between photons at inputs $\{i,j\}$.}

{We can obtain an independent estimate of this by comparing experiment to a theoretical model, where $\mathcal{I}_{i,j}$ is allowed to vary, and then minimize their variation distance $d{=}1{/}2\sum_k\left|p^{(2),\text{exp}}_k-p^{(2),\text{th}}_k\right|$ between experimental and theoretical distributions.} As described in the main text, $p^{(2),\text{exp}}$ is normalised to $\sum_{k}p^{(2),\text{th}}_k$. This being relevant when computing $d$ as distributions for different degrees of indistinguishability have different normalisation factors. For a given $2$-photon input $\{i,j\}$, $\mathcal{I}_{i,j}$ is taken as that at the global minimum in $d$. We obtain $\mathcal{I}_{1,2}{=}0.520$, $\mathcal{I}_{2,3}{=}0.540$, and $\mathcal{I}_{1,3}{=}0.643$, see Fig.~\ref{fig:SM_BS2}.

We carried out time-correlated measurements of two-photon interference on a $2\times2$ beam-splitter to independently verify these degrees of indistinguishability. It has been shown in ref.~\cite{loredo2016scalable} that the indistinguishability of two photons emitted by a semiconductor quantum dot depends on their emission temporal distance $\Delta t_e$. When both photons are emitted with the same polarisation from the quantum dot, their indistinguishability decreases monotonically in $\Delta t_e$. In our case we obtain $\mathcal{I}^{bs}_{12.5\text{ns}}{=}0.6360{\pm}0.0063$ for photons emitted with $\Delta t_e{=}12.5$~ns, and $\mathcal{I}^{bs}_{25\text{ns}}{=}0.6252{\pm}0.0065$ for $\Delta t_e{=}25$~ns, see Fig.~\ref{fig:SM_BS3}. Note that for these measurements photons are emitted with the same polarisation from the quantum dot.

The amount of indistinguishability $\mathcal{I}_{1,3}{=}0.643$, involving photons emitted with $\Delta t_e{=}25$~ns, and $\mathcal{I}^{bs}_{25\text{ns}}{=}0.6252{\pm}0.0065$ are in good agreement. Both $\mathcal{I}_{1,2}{=}0.520$, and $\mathcal{I}_{2,3}{=}0.540$ involve photons emitted with $\Delta t_e{=}12.5$~ns, therefore the minimisation method finds similar values, these however present {some discrepancy} with $\mathcal{I}^{bs}_{12.5\text{ns}}{=}0.6360{\pm}0.0063$. The quantum dot presents a small fine structure splitting of the exciton line, which in turn reduces the indistinguishability of photons emitted from two orthogonal emissions. Inputs $\{1,2\}$, and $\{2,3\}$ in the \textsc{BosonSampling} experiment contain photons separated by the first polarising beam-splitter in the source demultiplexer (see main text), thus they are emitted with orthogonal polarisations from the quantum dot and exhibit a reduced value of indistinguishability compared to photons emitted with the same polarisation, consistent with the obtained values.

\subsection{VI.~Three-photon interference}
We employ the theoretical model introduced in ref.~\cite{GenMPI:Walther} to describe the interference of $3$ photons, labeled $1$, $2$, and $3$, scattered across a linear network $\mathcal{L}$. In such case, the probability of detecting a $3$-fold coincidence at the output $\{o_1,o_2,o_3\}$ of $\mathcal{L}$ is:
	\begin{widetext}
	\begin{equation}\label{eqSM:p3}
		p^{(3)}=t_6^\dagger\left( \mathbb{I}+\rho_{1,2}\mathcal{I}_{1,2}+\rho_{2,3}\mathcal{I}_{2,3}+\rho_{1,3}\mathcal{I}_{1,3}+{\tilde\rho}\sqrt{\mathcal{I}_{1,2}}\sqrt{\mathcal{I}_{2,3}}\sqrt{\mathcal{I}_{1,3}} \right)t_6;
	\end{equation}
where
	\begin{equation}\label{eqSM:t6}\nonumber
		t_6=\begin{pmatrix}
\frac{1}{\sqrt6}\text{per}(\mathcal{T})\\ 
\frac{1}{\sqrt6}\text{det}(\mathcal{T})\\ 
\frac{1}{2\sqrt3}\text{imm}(\mathcal{T})+\frac{1}{2\sqrt3}\text{imm}(\mathcal{T}_{213})\\ 
\frac{1}{6}\text{imm}(\mathcal{T})-\frac{1}{3}\text{imm}(\mathcal{T}_{132})-\frac{1}{6}\text{imm}(\mathcal{T}_{213})+\frac{1}{3}\text{imm}(\mathcal{T}_{312})\\ 
\frac{1}{6}\text{imm}(\mathcal{T})+\frac{1}{3}\text{imm}(\mathcal{T}_{132})+\frac{1}{6}\text{imm}(\mathcal{T}_{213})+\frac{1}{3}\text{imm}(\mathcal{T}_{312})\\ 
-\frac{1}{2\sqrt3}\text{imm}(\mathcal{T})+\frac{1}{2\sqrt3}\text{imm}(\mathcal{T}_{213})
\end{pmatrix},
	\end{equation}
	
	\begin{equation}\nonumber
		\mathbb{I}=\begin{pmatrix}
1 & 0 & 0 & 0 & 0 &0 \\ 
 0& 1 &  0&  0&0  &0 \\ 
 0& 0 & 1 &  0&  0& 0\\ 
 0& 0 & 0 & 1 & 0 &0 \\ 
 0& 0 &  0&  0& 1 &0 \\ 
 0&0  &  0&0  & 0 &1 
\end{pmatrix},~~~~~
\rho_{1,2}=\begin{pmatrix}
1 & 0 & 0 & 0 & 0 &0 \\ 
 0& -1 &  0&  0&0  &0 \\ 
 0& 0 & 1 &  0&  0& 0\\ 
 0& 0 & 0 & -1 & 0 &0 \\ 
 0& 0 &  0&  0& 1 &0 \\ 
 0&0  &  0&0  & 0 &-1 
\end{pmatrix},~~~~~
		\rho_{2,3}=\begin{pmatrix}
1 & 0 & 0 & 0 & 0 &0 \\ 
 0& -1 &  0&  0&0  &0 \\ 
 0& 0 & -\frac{1}{2} &  -\frac{\sqrt3}{2}&  0& 0\\ 
 0& 0 & -\frac{\sqrt3}{2} & \frac{1}{2} & 0 &0 \\ 
 0& 0 &  0&  0& -\frac{1}{2} &  -\frac{\sqrt3}{2} \\ 
 0&0  &  0&0  & -\frac{\sqrt3}{2} & \frac{1}{2} 
\end{pmatrix},
	\end{equation}
	
	\begin{equation}\nonumber
		\rho_{1,3}=\begin{pmatrix}
1 & 0 & 0 & 0 & 0 &0 \\ 
 0& -1 &  0&  0&0  &0 \\ 
 0& 0 & -\frac{1}{2} &  \frac{\sqrt3}{2}&  0& 0\\ 
 0& 0 & \frac{\sqrt3}{2} & \frac{1}{2} & 0 &0 \\ 
 0& 0 &  0&  0& -\frac{1}{2} &  \frac{\sqrt3}{2} \\ 
 0&0  &  0&0  & \frac{\sqrt3}{2} & \frac{1}{2} 
\end{pmatrix},~~~~~~~~~~
\tilde{\rho}=\begin{pmatrix}
2 & 0 & 0 & 0 & 0 &0 \\ 
 0& 2 &  0&  0&0  &0 \\ 
 0& 0 & -1 &  0&  0& 0\\ 
 0& 0 & 0 & -1 & 0 &0 \\ 
 0& 0 &  0&  0& -1 &0 \\ 
 0&0  &  0&0  & 0 &-1 
\end{pmatrix},
	\end{equation}	
	\end{widetext}
$\mathcal{I}_{1,2}$, $\mathcal{I}_{2,3}$, and $\mathcal{I}_{1,3}$ are pair-wise indistinguishability values; $\mathcal{T}$ is a $3\times3$ submatrix built with rows 1, 2, and 3, and columns $o_1$, $o_2$, and $o_3$ of $\mathcal{L}$; $\mathcal{T}_{a,b,c}$ is the matrix $\mathcal{T}$ with rows 1, 2, and 3 rearranged in order $o_1$, $o_2$, and $o_3$; and the permanent (per), determinant (det), and immanant (imm) of a $3\times3$ matrix are defined as:
	\begin{eqnarray}\nonumber
		\text{per}\begin{pmatrix}
a & b & c \\ 
d & e & f\\ 
g & h & i
\end{pmatrix}&=&aei+bfg+cdh+ceg+bdi+afh,\\ \nonumber
	\text{det}\begin{pmatrix}
a & b & c \\ 
d & e & f\\ 
g & h & i
\end{pmatrix}&=&aei+bfg+cdh-ceg-bdi-afh,\\ \nonumber
	\text{imm}\begin{pmatrix}
a & b & c \\ 
d & e & f\\ 
g & h & i
\end{pmatrix}&=&2aei-bfg-cdh.
	\end{eqnarray}		

\subsection{VII.~Validation of \textsc{BosonSampling}}
Aaronson and Arkhipov proposed a protocol to test data against the uniform sampler~\cite{AAValidation}, as a counterargument to the claim~\cite{UnifDist:Eisert} that a large-scale \textsc{BosonSampling} implementation would fail to distinguish the experimental data even from that of the trivial one. The method---used in Fig.~3b in the main text---exploits available information of the sampling device---the transfer matrix $\mathcal{L}$---to define an estimator $P_\text{est}{=}\prod^n_{i{=}1}\sum^n_{j{=}1}\left|\overline{\mathcal{L}}_{i,j}\right|^2$, with $\overline{\mathcal{L}}_{i,j}$ the $n{\times}n$ submatrix of the $m{\times}m$ transfer matrix in an experiment involving $n$ bosons in $m$ modes. Unlike the permanent, $P_\text{est}$ is efficiently computable---thus, the protocol is scalable---and yet is correlated with the Boson Sampler probabilities. For the uniform distribution, the probability of one photon entering $\mathcal{L}$ in input $i$ and exiting in output $j$ is a constant (uniform) value $\left|\overline{\mathcal{L}}_{i,j}\right|^2{=}1{/}m$ across any input/output setting, thus the estimator takes the form $P_\text{est}^{\text{u}}{=}\left(n{/}m\right)^n$. If the sampling device is functioning correctly, one expects to observe more probable events more often; thus the method simply consists of computing $P_\text{est}$ for every event observed, and keeping track of a counter that is increased in one unit if $P_\text{est}{>}P_\text{est}^{\text{u}}$, and decreased in one unit otherwise. A resulting positive counter then validates the \textsc{BosonSampling} experiment by rejecting the hypothesis that the data originates from the uniform sampler. Experimental evidence supporting that this method works, even with small data samples and experimental imperfections, was reported in refs.~\cite{BSValidate:Sciarrino,BSValidate:Obrien}.

{A different protocol, used in Fig.~3c in the main text, to test the data against a distinguishable sampler was proposed and demonstrated by Spagnolo et. al.~\cite{BSValidate:Sciarrino}. This method, based on the likelihood ratio test, computes the relative---i.e., normalised to the no-collision space---quantum and classical probabilities, $p^{Q}$ and $p^{C}$, for every observed output event; a counter is increased in one unit if $p^{Q}{>}p^{C}$, and decreased in one unit otherwise. At the end of an experimental run a positive counter validates a correct functioning of the \textsc{BosonSampling} machine by rejecting the distinguishable sampler hypothesis.}

\bibliographystyle{aip}

\end{document}